\documentclass[pre, superscriptaddress, twocolumn]{revtex4-2}

\usepackage{graphics}
\usepackage{graphicx}
\usepackage{bbding}
\usepackage{mathrsfs}
\usepackage{amsmath}
\usepackage[table,xcdraw]{xcolor}
\usepackage{color}
\usepackage{natbib}
\usepackage{bookmark}

\definecolor{lightblue}{HTML}{01abe9}
\definecolor{red}{HTML}{f54b1a}
\definecolor{darkblue}{HTML}{1b346c}

\usepackage{hyperref}
\hypersetup{
  colorlinks = true,
  linkcolor = red,
  citecolor = darkblue,
  urlcolor = lightblue
}

\begin{document}

\date{\today}

\title{Hierarchical Dynamic Routing in Complex Networks via Topologically-decoupled and Cooperative Reinforcement Learning Agents}

\author{Shiyuan Hu}
\email[]{shiyuan.hu@huawei.com}
\affiliation{Network Technology Lab, Huawei Technologies Co., Ltd., Beijing 100095, China}
\author{Shihan Xiao}
\email[]{xiaoshihan@huawei.com}
\affiliation{Network Technology Lab, Huawei Technologies Co., Ltd., Beijing 100095, China}

\begin{abstract}
The transport capacity of a communication network can be characterized by the transition from a free-flow state to a congested state. Here, we propose a dynamic routing strategy in complex networks based on hierarchical bypass selections. The routing decisions are made by the reinforcement learning agents implemented at selected nodes with high betweenness centrality. The learning processes of the agents are decoupled from each other due to the degeneracy of their bypasses. Through interactions mediated by the underlying traffic dynamics, the agents act cooperatively, and coherent actions arise spontaneously. With only a small number of agents, the transport capacities are significantly improved, including in real-world Internet networks at the router level and the autonomous system level. Our strategy is also resilient to link removals.

\end{abstract}

\maketitle
\section{Introduction}

Efficient and congestion-free transport is critical in many networks, such as the Internet, the power grid, and airport networks. These networks often show complex structures with important effects on various dynamic processes~\cite{Watts98, Barabasi99, Boccaletti06}. Typically in communication networks, there exists a critical packet generation rate, beyond which congestion occurs~\cite{Takayasu96, Ohira98, Arenas01, Zhao05, Ding20}, i.e., the number of buffered packets increases at some intermediate nodes, leading to performance degradation or even collapse. The onset of congestion defines a transport capacity for the network, which depends not only on the network structure but also on the routing strategies. 

The widely adopted shortest-path (SP) routing in many systems is to send packets from source to destination through paths with the fewest number of links. However, SP routing is inefficient and may lead to severe congestion due to jamming at nodes with high usage, especially in networks with heterogeneous degree distributions~\cite{Yan06, Martino09}. By following static paths with the least sum of node degrees, the transport capacity is significantly improved~\cite{Yan06}. A variety of other elements have also been considered in developing efficient routing strategies, including local~\cite{Echenique04, Echenique05, Wang06, Tang11} and global~\cite{Zhang07, Ling10, Tan13} traffic conditions, packet priorities~\cite{Kim09, Du13}, and network resource allocation~\cite{Wu13}.

With the rapid expansion of scale and explosive growth of traffic volumes in modern communication networks, traffic control becomes increasingly difficult and complex. Instead of pushing the limit of transport capacity, we focus on two other perspectives: dynamic adaptation and distributed control. Dynamic routing can achieve better load balance than static routing by self-adapting to diverse network conditions, and distributed control is more feasible and reactive than centralized control in large-scale systems~\cite{Antonelli13, Marden15}. As an unsupervised learning and control approach, reinforcement learning (RL) has been applied to many domains~\cite{Mnih15, Boutaba18, Garnier21, Degrave22} and also to dynamic routing problems (e.g.,~\cite{Boyan93} and see reviews in~\cite{Hasan15, Dai21}). Recent studies of distributed dynamic routing with RL mainly focus on hop-by-hop routing~\cite{Mukhutdinov19, Ding20, Kang21, You22}. To guarantee fast convergence and packet delivery without loops, various techniques are needed, such as node identification using some encoding schemes~\cite{Mukhutdinov19, You22}, link reversal~\cite{Gafni81, Xiao20}, and limiting routing changes to only a few hops (otherwise following SP)~\cite{Kang21}. 

Although the dynamic routing strategies considered in previous studies can adapt to time-changing traffic conditions, several questions that are both fundamentally and practically important remain largely unexplored. Instead of selecting the next hop at each step, how to limit the number of routing changes and utilize topological properties to bypass congested nodes? If the routing decisions are distributed and made at each node, how to achieve a global optimum coherently? Another desired feature is low reliance on the communications between nodes, since frequent communications may add additional loads to the network. 

In this work, we propose a novel hierarchical dynamic (HD) routing strategy in heterogeneous networks based on sets of bypasses around selected nodes with high betweenness centrality (BC). The bypass decisions are made online according to the real-time traffic conditions by the RL agents implemented at each selected node. Due to the degeneracy of their action spaces, the agents are decoupled from each other, avoiding complex competitive multi-agent settings~\cite{Zhang21}. Although seemingly independent, different RL nodes interact with each other through the underlying traffic dynamics, and coherent actions arise spontaneously across the agents. Our strategy outperforms the least-degree (LD) routing~\cite{Yan06} with larger capacity and lower travel time. Quite remarkably, even with a small number of agents, the transport capacity is increased significantly. Different from the routing strategy based on global traffic conditions~\cite{Zhang07, Ling10, Tan13}, the agents can converge and make optimal decisions based only on the local traffic conditions, which eliminates the cost of global communications across the network. We further demonstrate that HD routing is resilient to link removals.

\section{Simulation setup}
\begin{figure}[t]
\centering
\includegraphics[bb= 0 8 240 126, scale=1.0, draft=false]{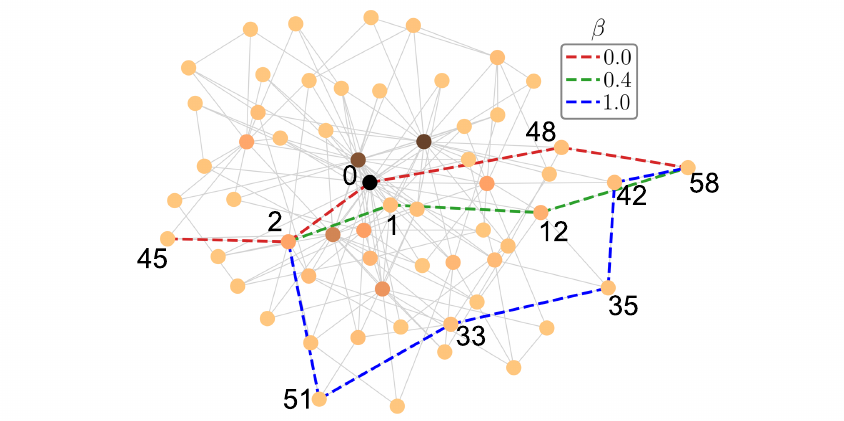}
\caption{Schematic of bypass in a BA network with 60 nodes. The nodes are colored according to their BC. Red line is the SP from node 45 to 58. Green and blue lines show two bypasses around node 0 computed using Eq.~(\ref{eq1}) with $\beta=0.4$ and $1.0$, respectively.}
\label{fig1}
\end{figure}

\footnotetext[1]{In Ref.~\cite{Yan06}, the static least-degree routing is computed with node degrees by setting $\beta=1$.}
\footnotetext[2]{We have also considered to include a bounded history of the queue lengths and found that its effect on the convergence of the learning process is negligible.}

\textit{Traffic model}.---Consider the transport of information packets on a complex network with $N$ nodes. Each node acts as both router and host~\cite{Zhao05, Yan06}. The traffic dynamics evolves forward with discrete time steps. At each time step, $R$ packets are generated on the network with random sources and destinations. Each node has a finite buffer (maximum queue length) and can forward at most one packet following first-in-first-out rule at each step. Additional arriving packets are dropped when the buffer is full. Delivered packets and dropped packets are both removed from the network.

We select nodes with high BC to implement the RL agents (hereinafter referred to as RL nodes), since they are more exposed to traffic under SP routing. The BC of node $\nu$ is defined as $b(\nu) = \sum_{s\neq d}\sigma_{s,d}(\nu)/\sigma_{s,d}$~\cite{Freeman77}, where $\sigma_{s,d}$ is the number of SPs from node $s$ to $d$ and $\sigma_{s,d}(\nu)$ is how many of them passing through $\nu$. Denote the set of RL nodes as $\mathcal{N}^{\alpha} = \{n_1^{\alpha}, \cdots, n_K^{\alpha}\}$, corresponding to the first $K$ ($K \ll N$) nodes with the highest BC. A packet is transported along one of the SPs after generation. If the next hop is one of the RL nodes, the packet may take a bypass around that RL node from its current location to its destination. We compute the bypass as follows. Consider a SP between source $s$ and destination $d$, $\{s, \cdots x, n_k^\alpha, \cdots d\}$ with $n_k^\alpha \in \mathcal{N}^{\alpha}$, and denote the set of all paths between $x$ and $d$ as $\mathcal{P}$, the bypass around $n_k^\alpha$ parameterized by $\beta$ is computed as
\begin{equation}\label{eq1}
p(x\to d; n_k^\alpha, \beta) = \mathop{\arg\min}_{p\in \mathcal{P}} \left(\sum_{\nu \in p} b(\nu)^{\beta}\right),
\end{equation}
where the summation is over nodes along the path $p\in \mathcal{P}$~\cite{[{The BC in Eq. (1) may be replaced with node degrees for faster computation, due to the correlations between degrees and BC, see }] [{}] Holme02, Note1}. A typical example of bypass in a Barab\'asi-Albert (BA) network~\cite{Barabasi99} is depicted in Fig.~\ref{fig1}. Equation~(\ref{eq1}) returns the SP for $\beta=0$ (no bypass). Given different values of $\beta$, a set of hierarchical bypasses can be obtained: as $\beta$ increases, the bypass becomes longer, but the average BC of nodes on the bypasses becomes smaller. 

Since Eq.~(\ref{eq1}) only considers topological information, the bypasses between any pair of nodes only need to form once. If a SP contains no RL node, the SP is followed without bypass; if a SP contains multiple RL nodes, the one with higher BC is bypassed, i.e., a packet bypasses at most once.

\textit{RL node}.---In RL, the intelligent agents interact with the environment and discover optimal policies through trial and error. At each RL node, we implement a Q-network~\cite{Mnih15}, a popular RL algorithm that takes continuous state input and approximates the action-value function corresponding to the optimal policy. Based on the state of the local traffic condition, the agent performs an $\epsilon-$greedy action selection at each time step and receives a reward, the traffic condition then moves to a new state. The experience is then stored, and the Q-network is updated by replaying the stored experiences.

We use a simple fully connected neural network of three layers with a rectifier nonlinearity~\cite{Nair10}. The received state of a RL node at time $t$ consists of its queue length and the queue lengths of other RL nodes~\cite{Note2}. The latter is considered since a packet may be directed to another RL node, and the traffic conditions there may affect the decision-making. The actions are the bypasses corresponding to a set of $\beta$ values. For all the RL nodes, we use the same set of $\beta$ values, $\{0, 0.2, 0.4, 0.6, 0.8, 1.0, 2.0\}$. The last ingredient is the reward function, which is computed as $\text{reward} = -\langle \text{travel time} \rangle - \langle \text{drop rate} \rangle$, where the first term is the travel time scaled by the maximum travel time following SP (the product of SP length and the buffer size) and the second term is the packet drop rate. Both $\langle \cdots \rangle$ denote average across packets. These two terms represent two competing effects: small travel time favors shorter paths, while low drop rate favors longer but less-congested paths with a lower risk of being dropped at severely congested nodes. Because both statistics are not immediately available after each action, we divide time into consecutive monitor intervals (MI)~\cite{Dong15, Jay19}. The RL nodes take actions at the beginning of each MI and perform the same actions throughout the MI. A RL node only records the packets directed by itself and computes the reward at the end of the MI. 

The training process breaks down into episodes. Each episode restarts the network traffic and ends after a fixed number of MIs. Throughout the simulations, the buffer size is 40, $\text{MI} =$ 5--15, and each episode has 50 MIs. The number of neurons in each hidden layer varies from 16--128, depending on the input size. 

\textit{Network topologies}.---Several practically relevant network topologies are used in our simulations. We first consider the BA scale-free network, which has a power-law degree distribution, $\phi(\kappa) \sim \kappa^{-3.0}$. To evaluate the effect of varying degree heterogeneity, we generate networks using configuration model~\cite{Newman03} with a $J$-tier composite exponential (CE-$J$) degree distribution, which is given by a sum of exponential distributions with mean $\lambda^{j+1}$ weighted by $a^{-(j+1)}$~\cite{Reynolds14}: $\phi(\kappa) \sim \sum_{j=0}^{J} a^{-(j+1)}\lambda^{j+1}\exp(-\lambda^{j+1}\kappa)$. When $J \to \infty$, $\phi(\kappa)$ resembles a power law; when $J\to 0$, $\phi(\kappa)$ degenerates into an exponential distribution. We also consider two real-world networks, the Internet at AS level~\cite{snapnets} and the router-level topology of an Internet service provider (ISP)~\cite{Spring02}. The AS network also has a power-law degree distribution, $\phi(\kappa)\sim \kappa^{-2.2}$. The basic topological features of these networks are summarized in Table~\ref{table1}. Both BA and AS networks have high degree heterogeneity.
\begin{table}[h]
\centering
\begin{tabular}{ c c c c c c } 
\hline
 & $N$ & $\langle \kappa\rangle$ & $\langle l\rangle$ & $\text{RSD}(\kappa)$ & $H$ \\
\hline 
BA & $10^3$ & 6.0 & 3.5 & 1.2 & 3.5 \\ 
CE-7 & $10^3$ & 5.8 & 3.8 & 0.8 & 2.4 \\
CE-3 &  $10^3$ & 5.4 & 4.2 & 0.6 & 1.8 \\
ISP & 624 & 17.0 & 3.4 & 1.2 & 2.5 \\
AS & 6474 & 3.9 & 3.7 & 6.4 & 42.4 \\
 \hline
\end{tabular}
\caption{Number of nodes $N$, mean degree $\langle \kappa \rangle$, average length of shortest paths $\langle l \rangle$, degree relative standard deviation $\text{RSD}(\kappa)$, and degree heterogeneity, $H=\langle \kappa^2\rangle/\langle \kappa\rangle^2$ of three synthetic networks and two real-world Internet networks used in our simulations. As a comparison, $\text{RSD}(\kappa)$ and $H$ for relatively homogeneous Erd\"os-R\'enyi network~\cite{Erdos60} with $N=1000$ and $\langle \kappa \rangle=6.0$ are 0.4 and 1.6, respectively. The statistics of the synthetic networks are averaged from 3 random generations.}
\label{table1}
\end{table}

\section{Results}

\begin{figure}[t]
\centering
\includegraphics[bb= 0 8 242 198, scale=1.0, draft=false]{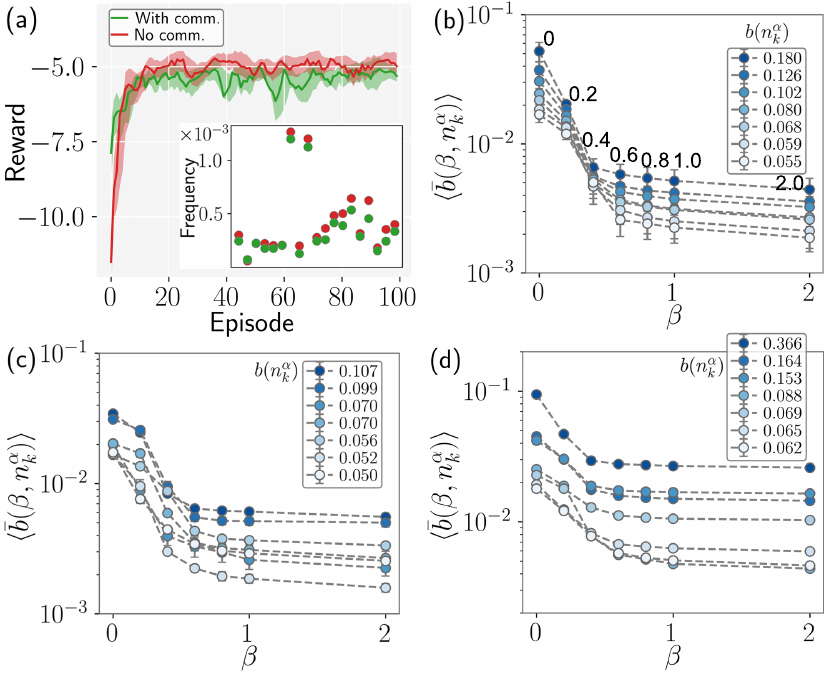}
\caption{The decoupling of the RL nodes. (a) Training reward averaged with a moving window of 5 episodes for the RL node with the highest BC in BA network with $K=20$. The green/red curve shows the reward obtained with/without communications to other RL nodes. The error bars are computed from simulations on three random network generations. Inset shows the frequency of directing packets to other RL nodes obtained from the last 30 episodes. (b)--(d) As a function of $\beta$, $\langle \bar{b}(\beta,n_k^{\alpha})\rangle$ of the first seven nodes with the highest BC (see legend) in (b) BA network, (c) ISP network, and (d) AS network. We normalize BC by $2/\left[(N-1)(N-2)\right]$.}
\label{fig2}
\end{figure}

To evaluate the effect of nonlocal traffic conditions, we implement 20 RL nodes in BA network. Figure~\ref{fig2}(a) shows that the communications between RL nodes on their traffic conditions has little effect on the reward, and both cases converge within 30 episodes, i.e., the learning process of a RL node only depends on its own traffic condition. This is because the frequency of redirecting packets to another RL node is around $1/N$ [Fig.~\ref{fig2}(a) inset], which is small in large networks. Similar results are observed in other networks explored in this study. We further compute the average BC of the bypasses, $\langle \bar{b}(\beta, n^{\alpha}_k)\rangle$, where $\bar{b}(\beta, n^{\alpha}_k)$ is the average BC of nodes on a bypass of $\beta$ around a RL node $n^{\alpha}_k$ and $\langle \cdots \rangle$ denotes average obtained by random sampling the source and destination pairs in the network. Figures~\ref{fig2}(b)--(d) demonstrate two folds of hierarchies of bypasses in BA, ISP, and AS networks. First, for the same RL node $\langle \bar{b}(\beta, n^{\alpha}_k) \rangle$ decreases as $\beta$ is increased and the bypass becomes more marginal. Second, for the same value of $\beta$, $\langle \bar{b}(\beta, n^{\alpha}_k) \rangle$ of a node with lower BC is, in general, degenerated to that of a node with higher BC, i.e., the action spaces of different RL nodes with different BC are decoupled from each other. Therefore, we speculate that the inter-node communications on their traffic conditions may not be important even when $K \sim N$. 

\begin{figure}[t]
\centering
\includegraphics[bb= 0 8 242 250, scale=1.0, draft=false]{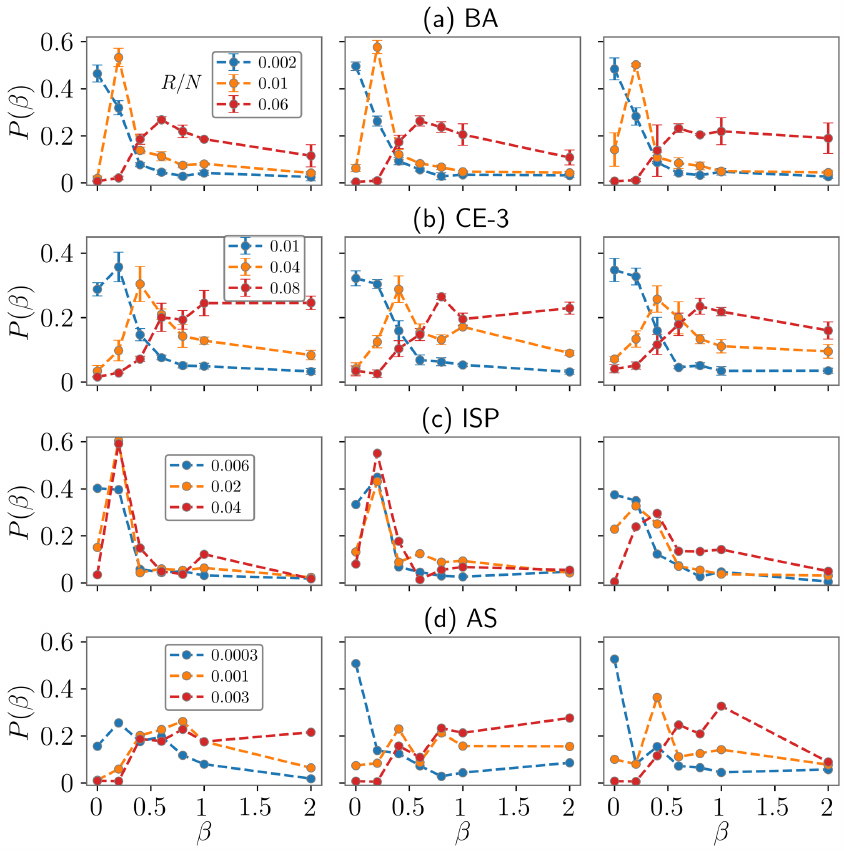}
\caption{The frequency distributions of actions $P(\beta)$ for three RL nodes (corresponding to three columns with decreasing BC from left to right) with different $R/N$ in (a) BA network, (b) CE-3 network, (c) ISP network, and (d) AS network. Results are obtained from the last 30 episodes.}
\label{fig3}
\end{figure}

We show the frequency distributions of actions (different values of $\beta$) in Fig.~\ref{fig3}. During the traffic simulation, even after convergence, the queue length at a RL node is time-changing, and different values of $\beta$ may be selected. However, the most frequent actions are different at different packet generation rates. At small $R/N$, packets travel across the network with little queuing, and the bypasses of small values of $\beta$ are frequently selected to minimize the travel time (blue lines in Fig.~\ref{fig3}). As $R/N$ increases, the RL nodes become more congested, and the frequencies of larger values of $\beta$ increase (orange and red lines). Arriving packets that need to wait in long queues or exceed the finite buffers are gradually redistributed to nodes with lower BC. 

\begin{figure}[b]
\centering
\includegraphics[bb= 0 10 242 275, scale=1.0, draft=false]{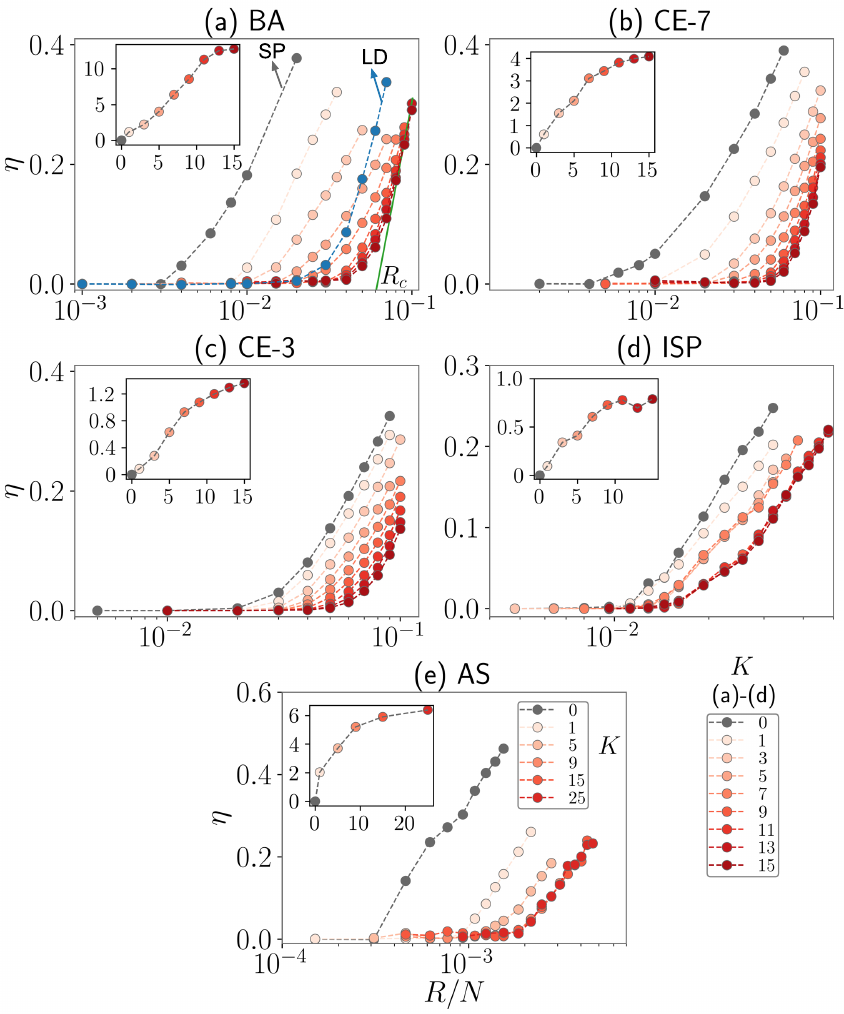}
\caption{The order parameter $\eta$ as a function of packet generation rate $R/N$ in semi-log scale for an increasing $K$ (from light red to dark red circles) in five different networks. The gray circles show data for SP routing and the blue circles show data for LD routing. For (a)-(d), $K=0$--15. The results on BA, CE-7, and CE-3 networks are averaged from independent simulations on three random network generations. The transport capacity $R_c$ is estimated as the intersection of the linear trend and $\eta=0$ (see green line as an example). Insets show the relative increase of $R_c$ as a function of $K$ compared with SP routing.}
\label{fig4}
\end{figure}
Another important observation from Fig.~\ref{fig3} is that, for different RL nodes in the same network, their frequency distributions are quite similar with the most frequent values of $\beta$ close to each other, even though the actions are taken independently based on their own traffic conditions. The coherent action selections indicate that different RL nodes act cooperatively and are non-competitive over network resources, since the bypasses of the same $\beta$ for different RL nodes are degenerated from each other [Figs.~\ref{fig2}(b)--(c)]. As an example, in BA network with $R/N=0.06$, the most frequency action for the RL node with the highest BC [Fig.~\ref{fig3}(a) left] is $\beta=0.6$. If the second RL node [Fig.~\ref{fig3}(a) middle] selects smaller values of $\beta$ more frequently, it would be discouraged by long queuing and even packet loss, since its bypasses of $\beta < 0.6$ have larger $\langle \bar{b}(\beta, n^{\alpha}_k) \rangle$ than the bypasses of $\beta=0.6$ of the first RL node [Fig.~\ref{fig2}(b)] and are already occupied by the packets directed from the first RL node. Actions of larger values of $\beta$ are also not optimal due to long bypass length. As a result, the most frequent action at $R/N=0.06$ for the second RL node is also around $\beta=0.6$. This hierarchical exploitation of network resources continues as we add more RL nodes. Therefore, the coherent actions depicted in Fig.~\ref{fig3} arise spontaneously without communications with each other, and the global optimum is achieved after each RL node behaves optimally. 

To characterize the transition from free-flow state to congested state, we compute the order parameter~\cite{Arenas01},
\begin{equation}\label{eq3}
\eta = \frac{1}{R}\frac{\langle W(t+\Delta t) - W(t)\rangle}{\Delta t}.
\end{equation}
Here, $\langle \cdots \rangle$ indicates a moving average over a time window of size $\Delta t$, and $W(t)$ is the sum of the number of in-transit and cumulative dropped packets at time $t$. At small $R/N$, packets are delivered with little queuing on the network and $\eta = 0$. The transport capacity is characterized by a critical generation rate $R_c/N$, beyond which packets start to accumulate in queues and $\eta > 0$. Figure~\ref{fig4} shows $\eta$ for five different networks with different number of RL nodes $K$. The increase of $R_c$ relative to SP routing is computed as, $\Delta R_c(K) = [R_c(K)-R_c(0)]/R_c(0)$ (Fig.~\ref{fig4} insets). We observe significant increases in $R_c$, especially in networks with high degree heterogeneity (BA and AS networks). 
\begin{figure}[t]
\centering
\includegraphics[bb= 0 8 242 176, scale=1.0, draft=false]{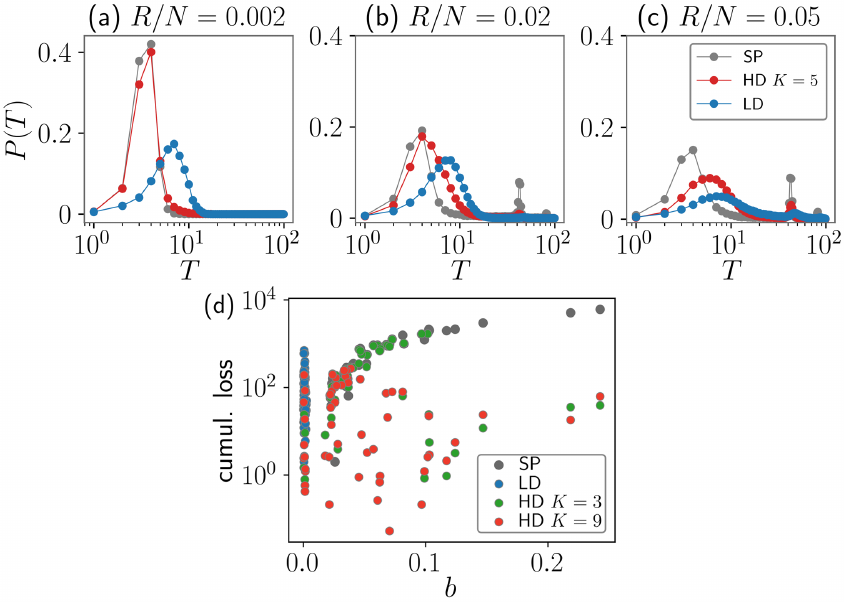}
\caption{(a)--(c) The frequency distributions of packet travel time $P(T)$ in semi-log scale for SP (gray circles), HD (red circles) with $K=5$, and LD routing (blue circles) at three different packet generation rates in BA network, (a) $R/N = 0.002$, (b) $R/N = 0.02$, and (c) $R/N = 0.05$. (d) The distribution of cumulative packet loss on BC at $t=500$.}
\label{fig5}
\end{figure}
In BA network, $R_c$ is increased by more than 10 times with $K/N \sim 0.01$; in AS network, $R_c$ is increased by around 6 times with $K/N \sim 0.003$. In these networks, as $K$ is increased, $R_c$ increases sharply first and then gradually saturates, since the routings of the majority of packets are controlled by a few RL nodes due to the fat-tailed $\phi(\kappa)$. As $\phi(\kappa)$ becomes more homogeneous from BA to CE-7, and then to CE-3, the increase of $R_c$ slows down. In CE-3 network, with $\langle H \rangle$ close to that of Erd\"os-R\'enyi network [Table~\ref{table1}], $R_c$ is still doubled with only 1\% RL nodes. Our strategy outperforms LD routing in BA network [Fig.~\ref{fig4}(a) blue line]. This is because in LD routing packets are distributed to nodes with small degrees and travel long distances, regardless of the congestion level. But in our strategy, a set of hierarchical bypasses of different lengths are dynamically selected depending on the traffic conditions [Fig.~\ref{fig3}].

To further demonstrate that our strategy achieves a balance between routing through hub nodes at the network center and nodes at the network periphery, we show in Fig.~\ref{fig5} the statistics of packet travel time and dropped packets. At small $R/N$, the frequency distribution of packet travel time of HD routing is close to that of SP routing, and both are much smaller than LD routing [Fig.~\ref{fig5}(a)]. As $R/N$ increases, the distribution of HD routing gradually shifts towards LD routing as more and more packets take longer bypasses. The peaks in SP routing in Figs.~\ref{fig5}(b) and \ref{fig5}(c) around $T = 40$ and 80 are due to overfilled buffers: packets at the end of a queue have to wait for the whole queue to be forwarded. This also indicates severe packet loss. Figure~\ref{fig5}(d) shows the distribution of cumulative packet loss on BC. As two extremes, there are considerable packet losses at nodes with high BC in SP routing and nodes with low BC in LD routing. The introduction of 3 RL nodes first significantly reduces packet losses by about two orders of magnitude at nodes with high BC [green circles in Figure~\ref{fig5}(d)]. As $K$ is increased from 3 to 9, the packet losses at nodes with intermediate BC around 0.05--0.1 are also reduced since more packets are redistributed to nodes with low BC.
\begin{figure}[b]
\centering
\includegraphics[bb= 0 8 245 100, scale=1.0, draft=false]{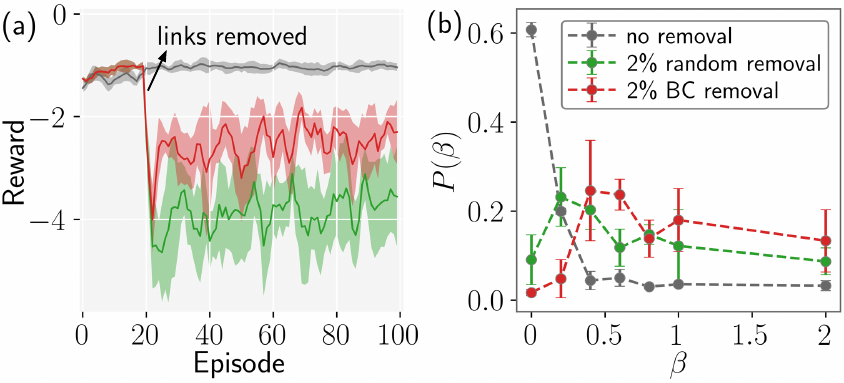}
\caption{Effect of link removals with $R/N=0.001$ and $K=3$ in BA network. (a) Training reward averaged with a moving window of 3 episodes for no removal (gray), random removal (green), and BC removal (red). (b) The frequency distributions of $\beta$ for the RL node with the highest BC. The error bars are computed from independent simulations on three random network generations.}
\label{fig6}
\end{figure}

Finally, we evaluate the effect of link removals. Two removal strategies are considered, random removal and removal with probabilities proportional to the average BC of the node pairs connecting to the links (BC removal). The former may be caused by attacks with little information about the network; the latter may be caused by attacks with partial information. Different from the previous approaches to increase the transport capacity by deliberate link removal or addition~\cite{Zhang2_07, Jiang11}, neither the RL nodes nor the packets are aware of the topological changes: the paths are not updated on the network after link removal. If the link between a packet's current node and its next-hop node is missing, the packet is dropped. We remove 2\% of the links at the end of episode 20. In the free-flow state ($R/N = 0.001$), Fig.~\ref{fig6}(a) shows that both removal strategies reduce the reward immediately caused by packet losses at missing links. Surprisingly, the recovery from BC removal is faster than random removal. This is due to the build-in hierarchies of the bypasses. Since at small $R/N$ most packets follow bypasses of small values of $\beta$, more penalty is assigned to these actions than actions of large values of $\beta$. Indeed, the frequencies of $\beta=0$ decrease, and the frequencies of larger values of $\beta$ increase for both removal strategies [Fig.~\ref{fig6}(b)]. However, for the BC removal, links in bypasses of small values of $\beta$ are removed with larger probabilities. Therefore, actions of small values of $\beta$ are penalized more compared with random removal, providing clearer feedback to the RL nodes that bypasses of small values of $\beta$ should be avoided. Indeed, the frequency distribution shifts towards larger $\beta$ values for the BC removal.

\section{Discussion}

We have studied a hierarchical dynamic routing strategy in heterogeneous complex networks with the routing decision-making distributed at topologically-decoupled RL nodes. The cooperative behaviors of the RL nodes and their coherent actions do not require explicit coordination through inter-node communications. It arises from the degeneracy of their actions spaces and indirect interactions mediated by the traffic dynamics. Most importantly, our results suggest that the transport capacity can be significantly increased by implementing only a small number of RL nodes, much smaller than the total number of nodes of the network. Our results may also be useful for the design of distributed intelligent agents in complex systems and provide insights into the understanding of their collective behaviors.

A balance of traffic between hub nodes and peripheral nodes has also been realized by computing an indicator of the traffic conditions at hub nodes, but the queue lengths at all nodes are required~\cite{Tan13}. Unlike hop-by-hop routing, our strategy is based on selections of bypasses between source and destination, and therefore loop-free routing is guaranteed. The conventional congestion control mechanism on the Internet along fixed paths typically only involves the source and destination nodes with the network structure functioning only to transport the packets. Based on the feedback information from the destination, the source can decrease its sending rate when congestion has likely occurred~\cite{Jacobson88}. In contrast to bypassing congested nodes dynamically, this end-to-end approach often limits network utilization. 

Our strategy may be further improved from several perspectives. For large-population agents in large-scale networks, we can speed up training by sharing the parameters of a single Q-network among all RL nodes~\cite{Zheng18, Mukhutdinov19}. The RL nodes can still make different decisions with their node identifications as additional inputs to the Q-network. We may also assign different weights to the two terms in the reward function and tune the preference over low travel time or low drop rate. In this study, we have only considered discrete action spaces (discrete $\beta$ values), continuous actions can be achieved with policy gradient algorithm~\cite{Lillicrap15}, but may require large networks to differentiate bypasses with small difference in $\beta$.

\begin{acknowledgments}
We thank Xiaofei Xu for fruitful discussions and Gang Yan for insightful comments and suggestions.
\end{acknowledgments}

\bibliographystyle{apsrev4-2}
\bibliography{reference}
\end{document}